%% file: Ostrogradsky.tex
\documentclass[a4paper,aps,onecolumn]{revtex4}
\usepackage{amsmath,amsfonts,amssymb,amscd,graphicx}
\RequirePackage[english]{babel}
\RequirePackage[latin1]{inputenc}
\RequirePackage[T1]{fontenc}

\newcommand{\be}{\begin{equation}}
\newcommand{\ee}{\end{equation}}
\newcommand{\bea}{\begin{eqnarray}}
\newcommand{\eea}{\end{eqnarray}}

\catcode`\@=12

\newfont{\gotico}{eufm10 scaled\magstephalf}
\newfont{\qvd}{msam10 scaled\magstephalf}

\def\de#1/de#2{\frac{\partial{#1}}{\partial{#2}}}
\def\d#1/d#2{\frac{d#1}{d#2}}
\def\sd#1/de#2/de#3{\ifx#2 \frac{\plus02\partial^{\@\@2}#1}{\plus90\partial\@#3^{\@2}}%
\else\frac{\plus02\partial^{\@\@2}#1}{\partial\?#2\@\partial\?#3}\fi}
\def\oD#1/d#2{\textstyle{\text{\large$\d{#1}/d{#2}$}}}
\def\De#1/de#2{\textstyle{\text{\large$\de{#1}/de{#2}$}}}
\def\SD#1/de#2/de#3{\ifx#2 \frac{\plus02\partial^{\@\@2}#1}
    {\plus90\partial\@#3^{\@2}} \else\frac{\plus02\partial^{\@\@2}#1}
    {\partial\@#2\@\partial\@#3}\fi}
\def\Sd#1/de#2/de#3{\textstyle{\text{\large$\sd{#1}/de{#2}/de{#3}$}}}
\def\SDs#1/de#2/de#3{\frac{\plus02\partial^{\@2}#1}
    {\partial\@#2\@\partial\@#3}}

\def\det{{\rm det}\,}
\def\rank{{\rm rank}\,}

\def\a{\alpha}
\def\b{\beta}
\def\g{\gamma}

\def\s{\sigma}
\def\th{\vartheta}

\def\w{\omega}

\def\A{\mathcal A}
\def\B{\mathcal B}

\def\F{\mathcal F}

\def\H{\mathcal H}
\def\I{\mathcal{I}\?}

\def\Re{\mathbb R}      \def\R{\Re}
\def\S{\mathcal S}
\def\Th{\Theta}
\def\V{{\mathcal V}_{n+1}}
\def\Q{\mathcal Q}

\def\jV{j_1(\V)}
\def\jjV{j_2(\V)}
\def\jQ{j_1(\Q)}
\def\CA{\mathcal C(\A)}
\def\VV{V(\V)}
\def\VVV{V^*(\V)}
\def\jQ{j_1(\Q)}

\def\@{\hskip.65pt}
\def\?{\hskip.3pt}
\def\plus#1#2{\vrule height#1pt width0pt depth#2pt}

\def\And{,\@\ldots\hskip-.4pt,}

\begin{document}
\title{\textbf{A new geometrical look at Ostrogradsky procedure.}}
\author{Enrico Massa$^1$\footnote{E-mail: massa@dime.unige.it}, Stefano Vignolo$^{1}$\footnote{E-mail: vignolo@dime.unige.it},
Roberto Cianci$^{1}$\footnote{E-mail: cianci@dime.unige.it}, Sante Carloni$^{2}$\footnote{E-mail: sante.carloni@tecnico.ulisboa.pt}}
\affiliation{$^{1}$DIME Sez. Metodi e Modelli Matematici, Universit\`{a} di Genova,\\
Piazzale Kennedy, Pad. D, 16129, Genova, ITALY\\
$^{2}$Centro Multidisciplinar de Astrofisica - CENTRA,
Instituto Superior Tecnico - IST,\\
Universidade de Lisboa - UL,
Avenida Rovisco Pais 1, 1049-001, Portugal.}
\date{\today}
\begin{abstract}
Making use of the modern techniques of non-holonomic geometry and constrained variational calculus, a revisitation of Ostrogradsky's Hamiltonian formulation of the
evolution equations determined by a Lagrangian of order $\@\ge 2\@$ in the derivatives of the configuration variables is presented.
\end{abstract}

\maketitle


\section{Introduction}
\input{Intro}

\section{Constrained variational calculus}\label{S2}
\input{par_2}

\section{The Ostrogradsky procedure revisited}\label{S3}
\input{par_3}

\subsection{Lagrangians of order $2$ in the derivatives}\label{S3A}
\input{par_3A}

\subsection{Lagrangians of order $N$ in the derivatives}\label{S3B}
\input{par_3B}

\bigskip
\bigskip
\noindent
{\bf Acknowledgments}

\bigskip
\noindent
SC  was supported by  the Funda\c{c}\~{a}o para a Ci\^{e}ncia e Tecnologia through project IF/00250/2013 and partly funded through H2020 ERC Consolidator Grant - Matter and strong-field gravity: New frontiers in Einstein's theory- grant agreement no. MaGRaTh-64659.

\input{ref}
\end{document}

%% file: Intro.tex
About twenty years after the first formulation of Hamiltonian mechanics, Ostrogradsky proposed a generalisation of Hamilton's procedure, valid for Lagrangians involving
derivatives of any order of the configuration variables \cite{O.,Woodard,Woodard:2006nt}.

In recent years, the study of this type of Lagrangians has been reconsidered in the context of gravitational physics and, in particular, in the development of a
theoretical framework for inflation and dark energy based on modifications of General Relativity (see e.g. \cite{HOGrev,Langlois:2015cwa,Chen:2012au}).

Despite this renewed interest, to the best of the authors' knowledge, a precise geometric interpretation of Ostrogradsky's construction is still missing. In an attempt
to fill this gap, we propose here a reformulation of Ostrogradsky's formalism in modern geometrical terms.

Given the event space, meant as a fibre bundle $\@\V\xrightarrow{t}\R\@$, we regard the $\@N^{th}\@$ jet bundle $\@j_N\/(\V)\@$ as an affine subbundle of the first jet
$\@j_1\/(j_{N-1}\/(\V))\@$ \cite{Pommaret,Saunders}. In this way, any problem involving a Lagrangian depending on the derivatives of order $\@\le N\@$ of the
configuration variables is converted into an ordinary \emph{constrained\/} variational problem.

The problem is then analysed, making use of a revisitation of Pontryagin's \emph{maximum principle\/} recently developed in \cite{Massa1} (in this connection, see also
\cite{Pontryagin,Giaquinta} and references therein). In the case of a non--degenerate Lagrangian $\@L\/(t,q^k,\dot q^k,\ddot q^k,\ldots\@)\@$, the algorithm picks out a
natural concept of ``phase space'', identifying it with a submanifold $\@\S\@$ of the contact bundle over $\@j_N\/(\V)\@$, uniquely determined by the Pontryagin
Hamiltonian associated with $\@L\@$.

In the resulting environment, the canonical momenta and the Ostrogradsky Hamiltonian are simply the pull--back of the coordinate functions along the fibres of the
contact bundle and of the Pontryagin Hamiltonian, while the Ostrogradsky equations reproduce the Hamilton--Pontryagin equations associated with the constrained
variational problem.

The layout of the paper is the following: in Section \ref{S2}, the geometrical setup for constrained variational calculus is briefly reviewed; Section \ref{S3} is then
devoted to the geometric reformulation of the Ostrogradsky procedure.

%% file: par_2.tex
In this section, we briefly review the geometrical formulation of constrained variational calculus along the lines described in \cite{Massa1}. The basic environment is a
$(n+1)$--dimensional fiber bundle $\@t:\V\to\Re\@$, referred to local fibred coordinates $\@t,q^1,\ldots,q^n\@$ and called the \emph{event space\/}. Every section
$\@\g:\Re\to\V\@$, locally described as $\@q^i=q^i\/(t)\@$, is interpreted as an evolution of an abstract system $\@\B\@$ with $n$ degrees of freedom: for instance, if
$\@\B\@$ represents a mechanical system, the manifold $\@\V\@$ is identified with the associated \emph{configuration space--time\/}, and the fibration $\@t:\V\to\Re\@$
with the absolute time function.

The first jet bundle $\@\jV\@$, referred to local jet coordinates $\@t,q^i,\dot q^i\@$, is called\vspace{-3pt} the \emph{velocity space}. Every section $\@\g:\Re\to\V\@$
admits a corresponding lift $\@j_1(\g):\Re\to\jV\@$, locally expressed as $\@q^i=q^i(t), \dot q^i=\oD q^i/dt\@$.\vspace{2pt}

The presence of non--holonomic constraints is geometrized through the assignment of a submanifold $i:\A\to\jV$ fibred over $\V$, as described by the commutative diagram
\begin{equation}\label{2.1}
\begin{CD}
\A     @>{i}>>  \jV             \\
@V{\pi} VV      @VV{\pi}V       \\
\V       @=         \V
\end{CD}
\end{equation}
all vertical arrows denoting bundle projections.
Referring $\@\A\@$ to local fibered coordinates $\@t,q^i,z^A\@$ ($A=1,\ldots,r<n$), the embedding $\@i:\A\to\jV\@$ is locally represented as
\begin{equation}\label{2.2}
\dot{q}^i = \psi^i(t,q^1,\ldots,q^n,z^1,\ldots,z^r)
\end{equation}
with $\@\rank \big\Vert \De\psi^i/de{z^A}\big\Vert=r\@$.

\smallskip
A section $\@\g:\Re\to\V\@$ is called \emph{admissible\/}\vspace{.6pt} if and only if there exists a section $\@\hat\g:\Re\to\A\@$ satisfying $\@j_1(\g)=i\cdot\hat\g\@$.
A section $\@\hat\g:\Re\to\A\@$\vspace{.6pt} is similarly called admissible if and only if $\@i\cdot\hat\g = j_1(\pi\cdot\hat\g)\@$. In coordinates, if $\@\hat\g\@$ is
described as $\@q^i=q^i(t), z^A=z^A(t)\@$, the admissibility condition is summarized into the system of first order ODE's
\begin{equation}\label{2.3}
\d q^i/dt= \psi^i(t,q^1(t),\ldots,q^n(t),z^1(t),\ldots,z^r(t))
\end{equation}

The geometry of the submanifold $\@\A\@$ has been extensively studied in the context of non--holonomic mechanics (see, among others, \cite{Massa2,Massa3} and references
therein). For the present purposes, we recall the concept of \emph{contact bundle} $\@\pi:\CA\to\A\@$, meant as the vector sub-bundle of the cotangent space
$\@T^*(\A)\@$ locally spanned by the contact $1$-forms
\begin{equation}\label{2.4}
\w^i := dq^i -\psi^i(t,q^k,z^A)\,dt
\end{equation}

Denoting by $\@\VV\subset T\/(\V)\@$ the vertical bundle relative to the fibration $t:\V\to\Re\@$\vspace{.6pt} and by $\@\VVV\@$ the associated dual bundle --- commonly
referred to as the \emph{phase space\/} --- the manifold $\@\CA\@$ is canonically isomorphic to the pull--back of $\@\VVV\@$ through the fibered morphism
\begin{equation}\label{2.5}
\begin{CD}
\CA         @>\pi>>         \VVV        \\
@V{\pi}VV               @VV{\pi}V       \\
\A          @>>\pi>         \V
\end{CD}
\end{equation}

We refer $\@\CA\@$ to fibred coordinates $\@t,q^i,z^A,p_i\@$, defined according to the identification $\@\s=p_i\/(\s)\@\w^i_{\;|\pi\/(\s)}$ $\forall\,\s\in\CA\@$.

An important geometrical attribute of the contact bundle is the its {\em Liouville $1$-form\/} $\@\Th\@$, locally expressed as
\footnote%
{For simplicity, we preserve the same notation for covariant objects on $\@\A\@$ and for their pull--back on $\@\CA\@$.}
\begin{equation}\label{2.6}
\Th := p_i\,\w^i = p_i\/\left(dq^i - \psi^i\/\left(t,q^k,z^A\right)\,dt\right)
\end{equation}

The geometrical framework outlined above provides the mathematical setting for an intrinsic formulation of constrained variational calculus. To this end, we consider an
action functional of the form
\begin{equation}\label{2.7}
\I\/[\g]:=\int_{\hat\g}\@L\,dt\@=\@\int^{t_1}_{t_0}L(t,q^i(t),z^A(t))\,dt
\end{equation}
assigning to each admissible section $\@\g:\Re\to\V\@$ a corresponding ``cost'', expressed as the integral of a \emph{Lagrangian function\/} $\@L(t,q^i,z^A)\in
\F\/(\A)\@$ along the lift $\@\hat\g:\Re\to\A\@$.
The aim is studying the (local) extremals of the functional \eqref{2.7} with respect to admissible deformations of $\@\g\@$ leaving the endpoints $\g\/(t_0),
\g\/(t_1)$ fixed.

This may be achieved observing that, under very general assumptions, the original problem is mathematically equivalent to a \emph{free\/} variational problem on the
contact bundle $\@\pi:\CA\to\A\@$.
The procedure, outlined in \cite{Massa1}, relies on the fact that, by means of the Liouville $1$-form \eqref{2.6}, every Lagrangian $\@L(t,q^i,z^A)\in \F\/(\A)\@$ may be
lifted to a $1$-form $\@\th_L\@$ over $\@\CA\@$ according to the prescription
\begin{equation}\label{2.8}
\th_L := L\,dt+\Th = (L-p_i\psi^i)\,dt + p_i\@dq^i :=-\@\H\,dt + p_i\@dq^i
\end{equation}

The function $\@\H\/(t,q^k,z^A,p_k)=-\@L\/(t,q^k,z^A)\@+\@p_i\,\psi^i(t,q^k,z^A)\@\in\F\/(\CA)\@$ is known in the literature as the \emph{Pontryagin
Hamiltonian}\vspace{1pt}. By means of the $1$--form \eqref{2.8}, to each section $\@\bar\g :[t_0,t_1]\to\CA\@$, expressed in coordinates as
$\@q^i=q^i\/(t)\@,\@z^A=z^A\/(t)\@,
\@p_i=p_i\/(t))$, we assign the action functional
\begin{equation}\label{2.9}
\bar \I\/[\bar\g]\@:=\int_{\bar\g}\@\th_L\@= \int_{t_0}^{t_1}\left[L(t,q^k(t),z^A(t)) +p_i(t)\left(\frac{dq^i}{dt}-\psi^i(t,q^k(t),z^A(t))\right)\right]dt
\end{equation}

The resulting setup is closely related to 
the original problem based on the functional \eqref{2.7} and on the constraints \eqref{2.3}. In fact, denoting by
$\@\nu:\CA\to\V\@$ the composite projection $\@\CA\to\A\to\V\@$, it turns out that every ``ordinary'' extremal of the original problem is the projection
$\@\g=\nu\cdot\bar\g\@$ of a solutions of the free variational problem based on the functional \eqref{2.9}
\footnote%
{For a precise definition of ordinariness see \cite{Massa1}.}.
More specifically, the requirement of stationarity of the action integral \eqref{2.9} under arbitrary deformations leaving the projections $\@\nu\/(\bar\g\/(t_0))\@,\,
\nu\/(\bar\g\/(t_1))$ fixed leads to $2n+r$ equations
\begin{subequations}\label{2.10}
\begin{align}
 & \d q^i/dt\,=\,\psi^i(t,q^k,z^A)\,=\,\de\H/de{p_i}                                    \label{2.10a}    \\[3pt]
 & \d p_i/dt\,=\,\de{L}/de{q^i} - p_k\@\de{\psi^k}/de{q^i}\,=\, -\@\de\H/de{q^i}        \label{2.10b}    \\[3pt]
 & p_i\@\de\psi^i/de{z^A} -\@\de L/de{z^A} \,=\, \de\H/de{z^A}\,=\,0                    \label{2.10c}
\end{align}
\end{subequations}
for the unknowns $\@q^i(t),z^A(t),p_i(t)\@$, identical to the \emph{Pontryagin equations\/} \cite{Pontryagin,Giaquinta} involved in the study of the constrained
functional \eqref{2.7}.

As far as the ordinary extremals are concerned, the original constrained variational problem in the event space is therefore equivalent to a free variational problem in
the contact bundle.

In order to analyse the content of the system \eqref{2.10}, it is convenient to start with eq.~\eqref{2.10c}. The latter identifies a subset of $\@\CA\@$, henceforth
denoted by $\@\S\@$.
%
%
%
The Hamiltonian $\@\H\@$ is called \emph{regular\/} if and only if the condition
\begin{equation}\label{2.11}
\det\biggl(\sd\H/de z^A/de{z^B}\biggr)_{\!\s}\ne\,0
\end{equation}
holds for all $\@\s\in\S\@$. When this is the case, eqs.~\eqref{2.10c} may be uniquely solved for the variables $z^A$, giving rise to a representation of the form
\begin{equation}\label{2.12}
z^A=z^A(t,q^i,p_i)
\end{equation}

Under the stated assumption, the subset $\@\S\@$ is therefore a $(2n+1)$--dimensional submanifold $\@i:\S\to\CA\@$, locally diffeomorphic~to the phase space $\@\VVV\@$.

The pull--back $\@H:=i^*(\H)\@$ of the Pontryagin Hamiltonian $\@\H\@$, expressed in coordinates as
\begin{equation}\label{2.13}
H(t,q^i,p_i)\@:=\@\H(t,q^i,z^A(t,q^k,p_k),p_i)\@=\@p_h\,\psi^h(t,q^i,z^A(t,q^k,p_k))\@-\@L(t,q^i,z^A(t,q^k,p_k))
\end{equation}
yields a proper Hamiltonian function on $\@\S\@$. Through the latter, the remaining equations \eqref{2.10} may be written as ordinary Hamilton equations. On account of
eqs.~\eqref{2.10c} we have in fact the identifications
\begin{subequations}\label{2.14}
\begin{align}
 & \de H/de{p_i}\,=\,\de\H/de{p_i}\,=\,\psi^i                                               \label{2.14a} \\[3pt]
 & \de H/de{q^i}\,=\,\de\H/de{q^i}\,=\,p_k\@\de\psi^k/de{q^i}\,-\,\de L/de{q^i}             \label{2.14b}
\end{align}
\end{subequations}
allowing to cast eqs.~\eqref{2.10a}, \eqref{2.10b} into the form
\begin{subequations}\label{2.15}
\begin{align}
& \d q^i/dt\,=\,\de H/de{p_i}                                       \label{2.15a} \\[3pt]
& \d p_i/dt\,=\,-\,\de H/de{q^i}                                    \label{2.15b}
\end{align}
\end{subequations}

The original constrained variational problem is thus reduced to a free Hamiltonian problem\vspace{1pt} in the submanifold $\@\S\@$, with Hamiltonian $\@H(t,q^i,p_i)\@$
identical to the pull--back $\@H=i^*(\H)\@$.

%% file: par_3.tex
In this section, we propose a revisitation of Ostrogradsky's construction of a Hamiltonian setup for the study of variational problems based on non--degenerate
Lagrangians $\@L\/(t,q^i,\dot{q}^i,\ddot q\@^i,\@\ldots\@)\@$ involving higher order derivatives of the configuration variables \cite{O.,Woodard}. The idea is regarding
any such $\@L\@$ as a function on a submanifold of a suitable velocity space, thereby reducing the original problem to a constrained one, of the kind described in the
Section \ref{S2}.

As we shall see, pursuing this viewpoint will provide an identification of the Ostrogradsky Hamiltonian with the pull--back \eqref{2.13} of the Pontryagin one, thus
opening the way to a self--consistent interpretation of Ostrogradsky's formalism in modern geometrical terms.

For the sake of simplicity, and to better fix the basic ideas and notations, we shall first consider Lagrangians of order $\@2\@$ in the derivatives. The procedure will
then be extended to higher order Lagrangians.

%% file: par_3A.tex
To start with, let us briefly review the Ostrogradsky procedure for Lagrangians of derivative order $2$. Given a Lagrangian of the form $L(t,q^i,\dot q\@^i,\ddot
q\@^i)$, the associated Euler--Lagrange equations read
\begin{equation}\label{3A.1}
\de L/de{q^i} - \d/dt\,\de L/de{\dot q\@^i}\,+\,\d^2/d{t^2}\,\de L/de{\ddot q\@^i}\,=\,0\@, \qquad i=1,\ldots,n
\end{equation}

Assuming the validity of non-degeneracy condition $\@\det\!\?\left\|\Sd L/de\ddot q\@^i/de{\ddot q\@^j}\right\|\ne0\@$, Ostrogradsky's idea consists  in adopting the
functions
\begin{equation}\label{3A.2}
q^i,\quad\dot q\@^i,\quad p^0_i:=\,\de L/de{\dot q\@^i}\,-\,\d/dt\,\de L/de{\ddot q\@^i}\,,\quad p^1_i:=\,\de L/de{\ddot q\@^i}
\end{equation}
as coordinates in a $(4n+1)$--dimensional \emph{phase space\/}, with $p^0_i$ and $p^1_i$ respectively meant as canonical momenta conjugate to the variables $q^i$ and
$\dot q\@^i$.

Under the stated non-degeneracy condition, the last set of equations \eqref{3A.2} can be solved for the unknowns $\@\ddot q\@^i$, giving rise to a representation of the
form
\begin{equation}\label{3A.3}
\ddot q\@^i =\@\ddot q\@^i\left(t,q^k,\dot q^k,p^1_k\right)
\end{equation}

In this way, introducing the \emph{Ostrogradsky Hamiltonian\/}
\begin{equation}\label{3A.4}
H:= p^0_i\@\dot q\@^i+\@p^1_i\@\ddot q\@^i-\@ L
\end{equation}
and expressing it in terms of the variables $\@q^i,\dot q\@^i, p^0_i\@,p^1_i\@$ through eqs.~\eqref{3A.3}, a straightforward calculation yields the relations
\be\nonumber
\de H/de{p^0_i} = \dot q\@^i, \qquad \de H/de{p^1_i} = \ddot q\@^i, \qquad \de H/de{q^i} =-\@\de L/de{q^i}\,,\qquad \de H/de{\dot q\@^i}\@=p^0_i\@-\@\de L/de{\dot q\@^i}
\end{equation}

In view of these, the content of eqs.~\eqref{3A.1}, \eqref{3A.2} may be cast into the Hamiltonian form
\begin{subequations}\label{3A.5}
\begin{alignat}{2}
 & \d q^i/dt\,=\@\dot q\@^i=\,\de H/de{p^0_i}\,, &&\d\dot q\@^i/dt\,=\@\ddot q\@^i=\,\de H/de{p^1_i}                                           \label{3A.5a} \\[3pt]
 & \d p^0_i/dt\,=\,\d/dt\,\de L/de{\dot q\@^i}\,-\,\d^2/d{t^2}\de L/de{\ddot q\@^i}\,=\,\de L/de{q^i}\,=\@-\@\de H/de{q^i}\,, \qquad\quad&&
 \d p^1_i/dt\,=\,\d/dt\,\de L/de{\ddot q\@^i}\,=\,\de L/de{\dot q\@^i}\,-\@p^0_i\,=\@-\@\de H/de{\dot q\@^i}                                     \label{3A.5b}
\end{alignat}
\end{subequations}

\vskip2pt
To clarify the geometrical meaning of the Ostrogradsky procedure, we focus on the fiber bundle $\@\V\to\Re\@$ and on the associated first jet bundle. For reasons that
will be clear soon, we change the notation $\jV$ into $\Q\@$, $q^i$ into~$q_0^i$, $\@\dot q\@^i\@$ into $\@q_1^i$, and regard the bundle $\@t:\Q\to\Re\@$, referred to
local coordinates $\@t,q_\a^i,\;\a=0\@,1\@$ as our new event space.\vspace{1pt}

By its very definition, the second jet bundle $\@\jjV\@$ is then (canonically isomorphic to) an affine subbundle of the first jet bundle $\@\jQ\@$, as expressed
by the commutative diagram 
\begin{equation}\label{3A.6}
\begin{CD}
\jjV        @>{i}>>    \jQ          \\
@V{\pi}VV              @VV{\pi}V    \\
\Q              @=     \Q
\end{CD}
\end{equation}

Referring $\jQ$ to jet coordinates $\@t,q_\a^i\@,\dot q_\a^i\@$, the image $\@i\?(\?\jjV)\subset\jQ\@$, henceforth denoted by $\@\A\@$, is locally described by the
equations $\@\dot q_0^i = q_1^i\@$. We can therefore refer $\@\A\@$ to local fibred coordinates $\@t,q_\a^i,z^i$, and represent the imbedding $\@\A\to\jQ\@$ through the
equations
(analogous to eqs.~\eqref{2.2}) 
\begin{equation}\label{3A.7}
\dot q^i_\a =\psi^i_\a\/(t,q_0^i,q_1^i,z^i)\@,
\end{equation}
with $\@\psi_0^i\/=q_1^i\@$ and $\@\psi_1^i\/=z^i\@$.

Alternatively, we may regard $\@\A\@$ as a fiber bundle over $\@\V\@$, related to $\@\jjV\@$ by the fibred isomorphism $\@(t,q_0^i,q_1^i,z^i)\longleftrightarrow
(t,q^i,\dot q\@^i,\ddot q\@^i)\@$.\vspace{1pt}

Collecting all results, we conclude that assigning a variational problem in $\@\V\,$, based on a Lagrangian $\@L(t,q^i,\dot q\@^i,\ddot
q\@^i)\in\F\/(\jjV)\@$\vspace{1pt}, is equivalent to assigning a constrained variational problem in $\@\Q\@$, with constraint submanifold $\@\A\to\jQ\@$ described by
eqs.~\eqref{3A.7} and Lagrangian $\@L(t,q_\a^i,z^i)\in\F\/(\A)\@$.

In the determination of the extremals, we can therefore proceed along the lines developed in Section \ref{S2}. To this end, we consider once again the contact bundle
$\CA$ over $\A$, and denote by $\@\nu:\CA\to\Q\@$ the composite projection $\@\CA\to\A\to\Q\@$, and by $t,q_\a^i,z^i,p^\a_i\@$ the local coordinates on $\@\CA\@$ defined
by the prescription
\begin{equation}\label{3A.8}
\s \@=\@ p^\a_i\/(\s)\,\big(dq_\a^i-\psi_\a^i\@dt\big)_{|\pi\/(\s)}   \qquad\forall\@\s\in \CA
\end{equation}
the summation convention being henceforth extended to all type of indices.

Starting with the Lagrangian $L(t,q_\a^i,z^i)\in\F\/(\A)$, we then construct the $1$-form
\begin{equation}\label{3A.10}
\th_L\@=\@L\@d\/t\@+\@p^\a_i\,\big(dq_\a^i-\psi_\a^i\@dt\big)\,=\,-\@\H\,dt\@+\@p^\a_i\,dq_\a^i\,\in\CA
\end{equation}
with
\begin{equation}\label{3A.9}
\H\/(t,q_\a^i,z^i,p^\a_i)\@=\@p^\a_i\@\psi_\a^i\@-\@L\/(t,q_\a^i,z^i)\@=\@ p^0_i\@q_1^i+\@p^1_i\@z^i-\@L\/(t,q_\a^i,z^i)
\end{equation}
denoting the Pontryagin Hamiltonian.
Eventually, we assign to each section $\@\bar\g :[t_0,t_1]\to\CA\@$ the action functional
\begin{equation}\label{3A.11}
\bar \I\/[\bar\g]\@:=\int_{\bar\g}\@\th_L\@= \int_{t_0}^{t_1}\biggl(-\@\H\/(t,q_\a^i,z^i,p^\a_i)\,+p^\a_i\,\d q_\a^i/dt\biggr)d\/t
\end{equation}

The request of stationarity of the integral \eqref{3A.11} under arbitrary deformations leaving the points $\@\nu\/(\bar\g\/(t_0))\@$, $\@\nu\/(\bar\g\/(t_1))\@$ fixed
leads to the Pontryagin equations
\begin{subequations}\label{3A.12}
\begin{align}
 & \d q_\a^{\@i}/dt\,=\,\de\H/de{p^\a_i}\,, \qquad  \d p^\a_i/dt\,=\@-\,\de\H/de{q_\a^{\@i}}                            \label{3A.12a}\\[3pt]
 & \de\H/de{z^i}\,=\,p^1_i\@-\@\de L/de{z^i}\,=\@0                                                                                                  \label{3A.12c}
\end{align}
\end{subequations}
for the unknowns $\@q_\a^{\@i}(t)\@,z^i(t),p^\a_i(t)\@$. These are the precise analogue of eqs.~\eqref{2.10} for the case in study.\vspace{1pt} In particular,
eqs.~\eqref{3A.12c} reproduce the content of the last set of eqs.~\eqref{3A.2} under the morphism $\@(t,q_0^i,q_1^i,z^i)\longleftrightarrow (t,q^i,\dot q\@^i,\ddot
q\@^i)\@$.\vspace{4pt}

Denoting by $\@\S\@$ the subset of $\@\CA\@$ described by eqs.~\eqref{3A.12c} and taking eq.~\eqref{3A.9} into account, it is readily seen that the non--degeneracy
condition $\@\det\!\?\left\|\Sd L/de\ddot q\@^i/de{\ddot q\@^j}\right\|\ne0\@$, here rephrased as $\@\det\!\?\left\|\Sd L/de z^i/de{z^j}\right\|\ne0\@$, automatically
ensures the regularity of the Pontryagin Hamiltonian \eqref{3A.9}.
We can therefore solve eqs.~\eqref{3A.12c} for the variables $z^i$, getting an expression of the form
\begin{equation}\label{3A.13}
z^i = z^i\@(t,\?q\@^k_\a\?,\@p^1_k)
\end{equation}
formally identical to eq.~\eqref{3A.3}

Exactly as it happened in Section \ref{S2}, eq.~\eqref{3A.13} allows to regard $\@\S\@$ as a submanifold $i:\S\to\CA$, locally diffeomorphic to the phase space
$V^*(\Q)$.
In view of eqs.~\eqref{3A.2}, \eqref{3A.12c}, the pull back of the Pontryagin Hamiltonian \eqref{3A.9} to the submanifold $\@\S\@$ yields the function
\begin{equation}\label{3A.14}
H\?(t,q_\a^i\@,\@p^\a_i)\,:=\@p^0_i\@q_1^i+\@p^1_i\@z^i\@(t,\?q\@^k_\a\?,\@p^1_k)\,-\,L\/(t,q_\a^i\@,z^i\/(t,\?q\@^k_\a\?,\@p^1_k))
\end{equation}
identical to the Ostrogradsky Hamiltonian \eqref{3A.4} and satisfying the relations
\begin{equation}\label{3A.15}
 \de H/de{p^\a_i}\,=\,\de\H/de{p^\a_i}\,, \qquad \de H/de{q_\a^{\@i}}\,=\,\de\H/de{q_\a^{\@i}}
\end{equation}

On account of the latter, eqs.~\eqref{3A.12a} may be cast into the canonical Hamiltonian form
\begin{subequations}\label{3A.16}
\begin{alignat}{2}
 & \d q_0^{\@i}/dt\,=\,\de H/de{p^0_i}\,,    \qquad &&\d q_1^{\@i}/dt\,=\,\de H/de{p^1_i}                       \label{3A.16a}\\[3pt]
 & \d p^0_i/dt\,=\@-\@\de H/de{q_0^{\@i}}\,, \qquad &&\d p^1_i/dt\,=\@-\@\de H/de{q_1^{\@i}}                    \label{3A.16b}
\end{alignat}
\end{subequations}
identical to the one taken by the Ostrogradsky equations \eqref{3A.5}.

%% file: par_3B.tex
The Ostrogradsky construction is easily extended to Lagrangians depending on higher order derivatives. To this end, let $\@j_N\/(\V)\@$ denote the $\@N^{th}\@$
jet--bundle of the event space, referred to fibred coordinates $\@t,q^i,q^i_1\And q^i_N\@$. Setting $\@q^i_0=q^i\@$, the Euler--Lagrange equations associated with a
Lagrangian $\@L\/(t,q^i,q^i_1\And q^i_N)\@$ are synthetically written as
\begin{equation}\label{3B.1}
\sum_{\a=0}^N\@(-1)^\a\@\d\@^\a/d{t^\a}\,\de L/de{q^i_\a}\,=0,                         \qquad i=1\And n
\end{equation}
For each $\@\a=0\And N-1\@$, let the canonical momentum $\@p^\a_i\@$ conjugate to the coordinate $\@q^i_\a\@$ be defined according to the prescription
\begin{equation}\label{3B.2}
p^\a_i := \sum_{\b=\a}^{N-1}\@(-1)^{\b-\a}\;\d\@^{\b-\a}/d{t^{\b-\a}}\,\de L/de{q^i_{\b+1}}
\end{equation}
whence, in particular
\begin{equation}\nonumber
p^{N-1}_i := \de L/de{q^i_N}\@(t,q^i,q^i_1\And q^i_N)               \tag{\ref{3B.2}'}
\end{equation}

The variables $\@t,q^i_\a\@,\@p^\a_i\,,\;\a=0\And N-1\@$ are regarded as coordinates in a $(2\?nN+1)$--dimensional \emph{phase space\/}\vspace{.4pt}. Under the
non-degeneracy assumption $\@\det\!\?\Big\|\Sd L/de q\@^i_N/de{q\@^j_N}\Big\|\ne0\@$\vspace{-2pt}, eqs.~(\ref{3B.2}') may be solved for $q^i_N$, giving rise to
expressions of the form
\begin{equation}\label{3B.3}
q^i_N = q^i_N\/(t,q^k_0\And q^k_{N-1},p_k^{N-1})
\end{equation}

The Ostrogradsky Hamiltonian is then defined as
\begin{equation}\label{3B.4}
H\/(t,q^i_0\And q^i_{N-1}\@,p_i^0\And p_i^{N-1})\@:=\sum_{\a=0}^{N-1}\@p_i^\a\@q^i_{\a+1}\@-\@L\/(t,q^i_0\And q^i_{N-1},q^i_N)
\end{equation}
with $\@q^i_N\@$ given by eq.~\eqref{3B.3}. In this way, taking eqs.~(\ref{3B.2}') into account, the Hamilton equations generated by the Hamiltonian \eqref{3B.4} are
easily recognized to be
\begin{subequations}\label{3B.5}
\begin{alignat}{2}
 & \d q^i_\a/dt\,=\,\de H/de{p_i^\a}\,=\@q^i_{\a+1} \quad\;(\a=0\And N-2), \qquad\quad
 &&\d q^i_{N-1}/dt\,=\,\de H/de{p_i^{N-1}}\,=\,q^i_N\/(t,q^k_0\And q^k_{N-1},p_k^{N-1})                             \label{3B.5a}\\[3pt]
 & \d p_i^0/dt\,=\@-\@\de H/de{q^i_0}\,=\,\de L/de{q^i_0}\,,\qquad\;
 &&\d p_i^\a/dt\@=-\@\de H/de{q^i_\a}\@=-\@p_i^{\a-1}+\@\de L/de{q^i_\a}\quad\;(\a=2\And N)         \label{3B.5b}
\end{alignat}
\end{subequations}

Conversely, a straightforward check shows that eqs.~\eqref{3B.5b}, together with eqs.~(\ref{3B.2}'), imply the validity of the Euler--Lagrange equations \eqref{3B.1}.

A deeper insight into the geometrical meaning of the Ostrogradsky algorithm is gained denoting by $\@Q:=j_{N-1}\/(\V)\@$ the $\@(N\!-\!1)^{th}$ jet bundle of the
fibration $\@t:\V\to\Re\@$, regarded as a fibre bundle $\@t:\Q\to\Re\@$, and by $\@\jQ\@$ the corresponding first jet bundle.
%
%
The $\@N^{th}$ jet bundle $\@j_N\/(\V)\@$ is then canonically isomorphic to an affine subbundle of $\@\jQ\@$, as summarized into the commutative diagram
\begin{equation*}
\begin{CD}
j_N\/(\V)               @>{i}>>    \jQ          \\
@V{\pi}VV                          @VV{\pi}V    \\
\Q                         @=      \Q
\end{CD}
\end{equation*}

Adopting $\@t,q^i_\a\@,\;(\a=0\And N-1)\@$ as local coordinates in $\@\Q\@$, and referring $\@\jQ\@$ to jet coordinates $\@t,q^i_\a,\dot q^i_\a\@$, the submanifold
$\@\A:=i\?(\?j_N\/(\V))\subset\jQ\@$  is locally described by the equations $\@\dot q^i_\a=q^i_{\a+1}\@$, $\@\a=0\And N-2\@$.

We can therefore refer $\@\A\@$ to local fibred coordinates $\@t,q^i_0\And q^i_{N-1},z^i\@$, and represent the imbedding $\@\A\to\jQ\@$ through the equations
\begin{equation}\label{3B.6}
\dot q^i_\a =\psi^i_\a\/(t,q_0^i\And q_{N-1}^i,z^i)\,,\qquad\quad \a=0\And N-1
\end{equation}
with $\@\psi^i_\a\/=q^i_{\a+1}\@$, $\,\a=0\And N-2\@$, and $\,\psi^i_{N-1}\/=z^i\@$.
Alternatively, we may regard $\@\A\@$ as a fiber bundle over $\@\V\@$, isomorphic to $\@j_N\/(\V)\@$ through the fibred morphism $\@(t,q^i_0\And q^i_{N-1},z^i)
\longleftrightarrow (t,q^i_0\And q^i_{N-1}\@,q^i_N)\@$.\vspace{1pt}

Once again, collecting all results, we conclude that assigning a variational problem in $\@\V\@$, based on a Lagrangian $\@L\/(t,q^i_0,\And
q^i_N)\in\F\/(\?j_N\/(\V))\@$, is equivalent to assigning a constrained variational problem in $\@\Q\@$, with constraint submanifold $\@\A\to\jQ\@$ described by
eqs.~\eqref{3B.6} and Lagrangian $\@L\/(t,q^i_0\And q^i_{N-1},z^i)\in\F\/(\A)\@$.

The constrained problem in $\@\Q\@$ may then be lifted to a free variational problem on the contact bundle $\CA$, referred to fibred coordinates $\@t,q^i_\a,z^i,
p_i^\a\@$, $\,\a=0\And N-1\@$. The procedure, identical to the one exploited in Sect.~\ref{S3A}, culminates in the introduction of the Pontryagin Hamiltonian
\begin{equation}\label{3B.7}
\H\/(t,q^i_\a,z^i,p_i^\a)\@=\@\sum_{\a=0}^{N-1}\@p_i^{\a}\@\psi^i_\a\@-\@L\/(t,q^i_\a,z^i)\@=
\sum_{\a=0}^{N-2}\@p_i^\a\@q^i_{\a+1}\,+\,p_i^{N-1}\@z^i\,-\,L\/(t,q^i\And q^i_{N-1},z^i)
\end{equation}

Preserving the notation $\@\CA\xrightarrow{\nu}\Q\@$ for the composite map $\@\CA\to\A\to\Q\@$, to each section $\@\bar\g:[t_0,t_1]\to\CA\@$ we now assign the action
functional
\begin{equation}\label{3B.9}
\bar \I\/[\bar\g]\@:=\int_{\bar\g}\@-\@\H\,d\/t\,+\,\sum_{\a=0}^{N-1}\@p_i^{\a}\,dq^i_\a\@=
\int_{t_0}^{t_1}\biggl(-\@\H\@+\sum_{\a=0}^{N-1}\@p_i^{\a}\,\d q^i_\a/dt\biggr)d\/t
\end{equation}

Imposing stationarity of the latter under arbitrary deformations leaving the projections $\@\nu\/(\bar\g\/(t_0))\@$, $\@\nu\/(\bar\g\/(t_1))\@$ fixed leads to the
Pontryagin equations
\begin{subequations}\label{3B.10}
\begin{align}
 & \d q_\a^{\@i}/dt\,=\,\de\H/de{p^\a_i}\,, \qquad  \d p^\a_i/dt\,=\@-\,\de\H/de{q_\a^{\@i}}                            \label{3B.10a}\\[3pt]
 & \de\H/de{z^i}\,=\,p_i^{N-1}\@-\,\de L/de{z^i}\,=\@0                                                                  \label{3B.10b}
\end{align}
\end{subequations}
for the unknowns $\@q_\a^{\@i}(t)\@,z^i(t),p^\a_i(t)\@$\vspace{1pt}. Eqs.~\eqref{3B.10b} reproduce the content of eqs.~(\ref{3B.2}') under the morphism $\@(t,q^i_0\And
q^i_{N-1},z^i)\longleftrightarrow (t,q^i_0\And q^i_{N-1}\@,q^i_N)\@$.\vspace{1pt}

Denoting by $\@\S\@$ the subset of $\@\CA\@$ described by eqs.~\eqref{3B.10b}, it is readily seen that the non--degeneracy condition $\Big\|\Sd L/de
q^i_N/de{q^j_N}\Big\|\ne0\@$, here rephrased as $\@\det\!\?\Big\|\Sd L/de z^i/de{z^j}\Big\|\ne0\@$\vspace{1pt}, automatically ensures the regularity of the Pontryagin
Hamiltonian \eqref{3B.7}.
We can therefore solve eqs.~\eqref{3B.10b} for the variables $z^i$, getting the expression
\begin{equation}\label{3B.11}
z^i\@=\@z^i\,(t,q^k_0\And q^k_{N-1},p_k^{N-1})
\end{equation}
formally identical to eq.~\eqref{3B.3}.

Eqs.~\eqref{3B.11} point out that the subset $\@\S\subset \CA\@$ is in fact a \emph{submanifold\/} $i:\S\to\CA$, locally diffeomorphic to the phase space $V^*(Q)$. A
straightforward check shows that the pull back $H:=i^*(\mathcal H)$ of the Pontryagin Hamiltonian \eqref{3B.7} determines a proper Hamiltonian function on $V^*(Q)$,
identical to the Ostrogradsky Hamiltonian \eqref{3B.4}, and that the Hamilton equations generated by $H$ coincide with the Ostrogradsky equations \eqref{3B.5}.